\documentclass[final,5p,times,twocolumn]{elsarticle}

\usepackage{amssymb}

\journal{Astroparticle Physics}

\newcommand{\dbd}{0$\nu$DBD}
\newcommand{\TEO}{TeO$_2$}
\begin{document}

\begin{frontmatter}



\title{Model for the Cherenkov light emission of {\TEO} cryogenic calorimeters}

\author[label1,label2]{N.~Casali} \ead{nicola.casali@roma1.infn.it}
\address[label1]{Dipartimento di Fisica - Sapienza Universit\`{a} di Roma, Piazzale Aldo Moro 2, 00185, Roma - Italy}
\address[label2]{INFN - Sezione di Roma, Piazzale Aldo Moro 2, 00185, Roma - Italy}



\begin{abstract}
The most sensitive process able to probe the Majorana nature of neutrinos and discover Lepton Number Violation is the neutrino-less double beta decay. Thanks to the excellent energy resolution, efficiency and intrinsic radio-purity, cryogenic calorimeters are primed for the search for this process. A novel approach able to improve the sensitivity of the current experiments is the rejection of $\alpha$ interactions, that represents the dominant background source. In {\TEO} calorimeters, $\alpha$ particles can be tagged as, in contrast to electrons, they do not emit Cherenkov light.
Nevertheless, the very low amount of detected Cherenkov light undermines the complete rejection of $\alpha$ background.\\
In this paper we compare the results obtained in previous measurements of the {\TEO} light yield with a detailed Monte Carlo simulation able to reproduce the number of Cherenkov photons produced in $\beta/\gamma$ interactions within the calorimeter and their propagation in the experimental set-up. 
We demonstrate that the light yield detectable from a $5\times5\times5$~cm$^{3}$ {\TEO} bolometer can be increased by up to 60\% by increasing the surface roughness of the crystal and improving the light detector design.\\
Moreover, we study the possibility to disentangle $\alpha$, $\beta$ and $\gamma$ interactions, which represent the ultimate background source. Unfortunately $\gamma$ rejection is not feasible but $\alpha$ rejection can be achieved exploiting high sensitivity light detectors.
\end{abstract}

\begin{keyword}
      Double beta decay \sep Neutrino mass and mixing \sep Bolometer \sep Cherenkov radiation \sep Monte Carlo methods


\end{keyword}

\end{frontmatter}


\section{Introduction}
\label{sec:introduction}

Neutrino-less double beta decay ({\dbd}) is an extremely rare process (if it occurs at all) in which a nucleus undergoes two simultaneous beta decays without emitting any neutrinos. Since the absence of emitted neutrinos would violate the lepton number conservation, the {\dbd} cannot be accommodated in the Standard Model. Therefore, the observation of this process would have several implications for particle physics, astrophysics and cosmology. The most common theoretical frameworks account for {\dbd} assuming that neutrinos are Majorana particles. This means that, in contrast to all the other known fermions, they must coincide with their own antiparticles. 
The measurement of the half-life of {\dbd} ($T^{0\nu}_{1/2}$) would allow to infer the effective Majorana neutrino mass ($m_{\beta\beta}$), provided the dependance: $T^{0\nu}_{1/2}\propto1/m^{2}_{\beta\beta}$. $m_{\beta\beta}$ is the coherent sum of the three neutrino mass eigenstates according to the PMNS neutrino mixing matrix and including two Majorana phases, thus it depends on the mass hierarchy of neutrino (see Refs~\cite{DellOro:2016tmg,Feruglio:2002af,Strumia:2005tc} and the references therein for a complete discussion on the relationship between {\dbd} and neutrino masses).

CUORE (\textit{Cryogenic Underground Observatory for Rare Events})~\cite{Artusa:2014lgv} is an array of 988 {\TEO} cryogenic calorimeters, historically also called bolometers, of $5\times5\times5$~cm$^{3}$ each. Starting operation in 2017 it will become one of the most sensitive experiments searching for {\dbd} of $^{130}$Te. The signal produced by this reaction consists of two electrons with a total kinetic energy of about 2.527~MeV (the Q-value of the transition)~\cite{Redshaw:2009cf}. Unfortunately the $\alpha$ background rate, at the level of $0.01$~counts/keV/kg/y~\cite{Aguirre:2014lua}, will limit the sensitivity to $T^{0\nu}_{1/2}$ and, consequently to the effective Majorana mass. The sensitivity to $m_{\beta\beta}$ will be at the level of $0.13-0.05$~eV in 5 years live time, i.e. the upper limit of the inverted hierarchy region of the neutrino masses~\cite{Artusa:2014lgv}. Next generation {\dbd} experiments, such as CUPID (\textit{Cuore Upgrade with Particle IDentification})~\cite{Wang:2015raa,Wang:2015taa}, aim to reach a sensitivity to $m_{\beta\beta}$ of $\sim0.01$~eV, i.e. the lower limit of the inverted hierarchy region. To reach this ambitious goal the source mass must be increased and the background in the region of interest dramatically reduced~\cite{Artusa:2014wnl}. The CUPID collaboration aims to increase the number of {\dbd} emitters, i.e. the source mass, using crystals grown from enriched material. The background suppression can be achieved by discriminating $\beta/\gamma$ against $\alpha$ events by means of the different light yield produced in the interactions within a scintillating bolometer~\cite{BolScintDBD}. The main candidates for this technique are ZnSe~\cite{Artusa:2016maw,Beeman:2013vda}, ZnMoO$_{4}$~\cite{Cardani:2013mja,Beeman:2012jd,Armengaud:2015hda} or Li$_{2}$MoO$_{4}$~\cite{Cardani:2013dia} scintillating crystals.

The same technique cannot be used for {\TEO} as this crystal does not scintillate~\cite{Bellini:2012rc,Casali:2013bva} and also no difference exists between the shape of the pulses produced by $\alpha$ and $\beta/\gamma$ interactions~\cite{Bellini:2010iw}.

However, the many advantages offered by this material in terms of bolometric performance and lower enrichment cost with respect to other candidate nuclei provided a strong motivation to pursue another, challenging, option: the active background rejection technique can be applied to the {\TEO} bolometer exploiting, instead of the scintillation light, the Cherenkov radiation. Indeed, as proposed in Ref.~\cite{TabarellideFatis:2009zz} detecting the Cherenkov radiation produced in the {\TEO} crystal only by electrons (see section~\ref{subsec:NumberOfCherenkovPhotons}), it is possible to disentangle the $\beta/\gamma$ interactions from the $\alpha$ interactions.

In order to measure the Cherenkov light yield (LY) of {\TEO} bolometers several tests coupling germanium light detectors~\cite{Beeman:2013zva} were performed in a dilution refrigerator working at 10~mK and located deep underground in the Hall C of Laboratori Nazionali del Gran Sasso. The first indication of the feasibility of this technique to reject the $\alpha$ background can be found in Ref.~\cite{Beeman:2011yc}. In this work a small {\TEO} crystal ($3.0\times2.4\times2.8$~cm$^3$) doped with natural Sm and covered with 3M VM2002 reflective foils was faced to a germanium light detector. The amount of Cherenkov light detected at {\dbd} energy of $^{130}$Te for $\beta/\gamma$ interactions was about 173~eV; no light was detected for $\alpha$ particle interactions. A more systematic study of the Cherenkov LY emitted by a CUORE-size {\TEO} crystal was done in Ref.~\cite{Casali:2014vvt} where several reflectors and detector configurations were tested in order to optimize the photon collection. The result of this study was that the maximum amount of Cherenkov light detected from a CUORE {\TEO} crystal is about 100~eV at the Q-value of $^{130}$Te {\dbd}, irrespective of the reflector (aluminum, PTFE tape, 3M VM2002). This, combined with the poor baseline resolution of the light detectors (80~eV RMS) prevented the complete $\alpha$ particles rejection. As discussed in Ref.~\cite{Casali:2014vvt} the reduced light collection efficiency and its invariance over the several reflectors tested suggests that most of the emitted light is reabsorbed by the {\TEO} crystal because of its optical properties.

The purpose of this paper is to demonstrate this hypothesis and to quantify its effects exploiting a detailed Monte Carlo simulation of the Cherenkov photon production and propagation in the crystal and in an experimental set-up similar to the one described in Ref.~\cite{Casali:2014vvt}. Thanks to the simulation, it is also possible to identify the parameters able to improve the light collection efficiency of the set-up. Finally, the simulation demonstrates that the Cherenkov radiation tagging technique does not allow for the discrimination between $\beta$ and $\gamma$ interactions, that will constitute the ultimate background source for the bolometric experiments based on {\TEO} crystals. Concerning the $\alpha$ background rejection, the simulation results confirm that it can be achieved exploiting a high sensitivity light detector.

\section{Cherenkov photon production in a {\TEO} crystal}
\label{sec:CherenkovTheory}


The number of Cherenkov photons produced per unit path length and per unit wavelength of a particle with charge $ze$ is~\cite{PhysRev.52.378}:
\begin{equation}
\frac{dN^2}{dxd\lambda}= \frac{2\pi \alpha z^{2}}{\lambda^{2}}\left( 1 - \frac{1}{\beta^{2}\left(n\left(\lambda\right)\right)^2}\right)\label{eq:NumberofPhotons}
\end{equation}
where $\alpha$ is the fine-structure constant, $\beta$ is the ratio between the velocity of the particle and the speed of light, $\lambda$ the wavelength of the Cherenkov photons, and $n(\lambda)$ the refractive index of the material. Eq.~\ref{eq:NumberofPhotons} is valid as long as it is positive, namely as long as the charged particle moves faster than light in the respective medium ($\beta n>1$). As shown in Eq.~\ref{eq:NumberofPhotons} the evaluation of the number of Cherenkov photons produced by a particle interaction within a {\TEO} crystal requires the knowledge of its optical properties and the particle range. Therefore in the following sections the optical properties of {\TEO} crystals will be presented together with its stopping power for electrons, in order to perform an estimation of the Cherenkov photons emitted in the $\beta/\gamma$ interactions.

\subsection{Optical properties of {\TEO} crystals}
\label{subsec:OpticalProperties}
\begin{figure}[htbp]
\includegraphics[width=.5\textwidth]{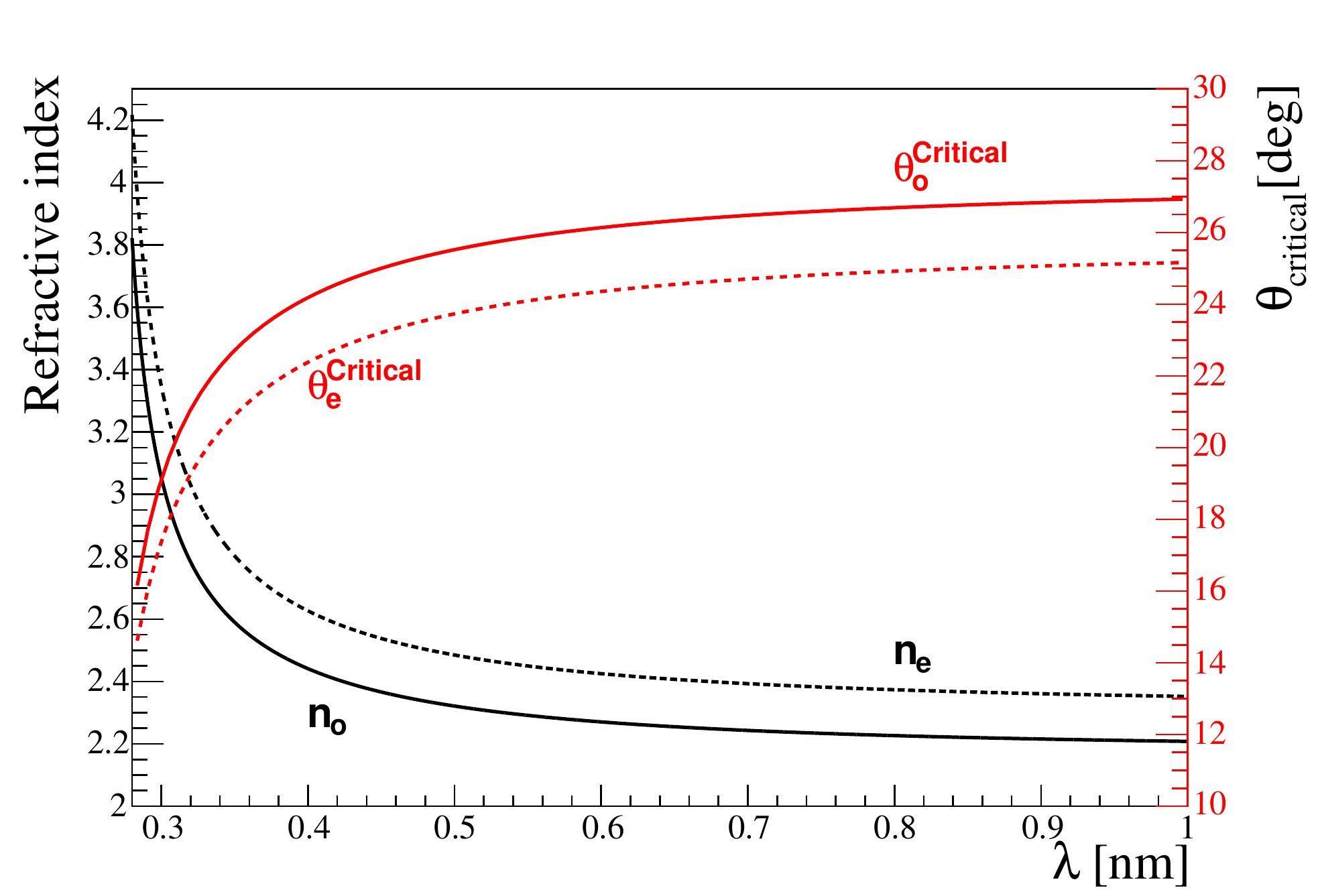}
\caption{Ordinary (n$_{\rm{o}}$, continuous black line) and extraordinary (n$_{\rm{e}}$, dotted black line) refractive indices of {\TEO} crystal as function of wavelength at room temperature~\cite{PhysRevB.4.3736}. $\theta^{Critical}$ is the critical angle for total internal reflection for an optical photon that hits the internal surface of a {\TEO} crystal surrounded by vacuum using the ordinary ($\theta^{Critical}_{\rm{o}}$, continuous red line) and extraordinary ($\theta^{Critical}_{\rm{e}}$, dotted red line) refractive indices.}\label{fig:RIndex}
\end{figure}

{\TEO} is a birefringent material. The ordinary and extraordinary refractive indices (n$_{\rm{o}}$ and n$_{\rm{e}}$, respectively) are shown in Fig.~\ref{fig:RIndex}.
Both the refractive indices are described by the Sellmeier equation~\cite{Sell}:
\begin{equation}
\label{eq:sellmeier}
n(\lambda)^2-1=\frac{A\lambda^2}{\lambda^2-\lambda_{1}^{2}}+\frac{B\lambda^2}{\lambda^2-\lambda_{2}^{2}}
\end{equation}
The Sellmeier parameters for the {\TEO} crystal are shown in Table~\ref{Tab:SellmeierParam}.
\begin{table}[htbp]
\begin{centering}
\begin{tabular}{ccccc}
\hline
& A & $\lambda_{1}$ [$\mu$m] & B & $\lambda_{2}$ [$\mu$m] \\
\hline
\hline
n$_{\rm{o}}$ & 2.584 & 0.1342 & 1.157 & 0.2638 \\
\hline
n$_{\rm{e}}$ & 2.823 & 0.1342 & 1.542 & 0.2631 \\
\hline
\end{tabular}
\caption{\small{Coefficients of the Sellmeier equations for the {\TEO} crystal at room temperature~\cite{PhysRevB.4.3736}.}}\label{Tab:SellmeierParam}
\end{centering}
\end{table}

As {\TEO} crystals are currently produced for acousto-optic devices, special attention is given in literature only to optical properties connected to this application. The study of optical characteristics as transmission and reflection near and above the fundamental absorption edge, at low temperatures, were carried out down to only 80~K \cite{doi:10.1143/JPSJ.48.505} and experimental results \cite{PhysRevB.4.3736,doi:10.1143/JPSJ.48.505} do not allow for a definitive interpretation of the electronic band structure for {\TEO} crystals. It is also missing from the literature a detailed study of the agreement between calculated electronic structure \cite{PhysRevB.69.193101} and optical transmission measurements.
\begin{figure}[htbp]
 \includegraphics[width=.5\textwidth]{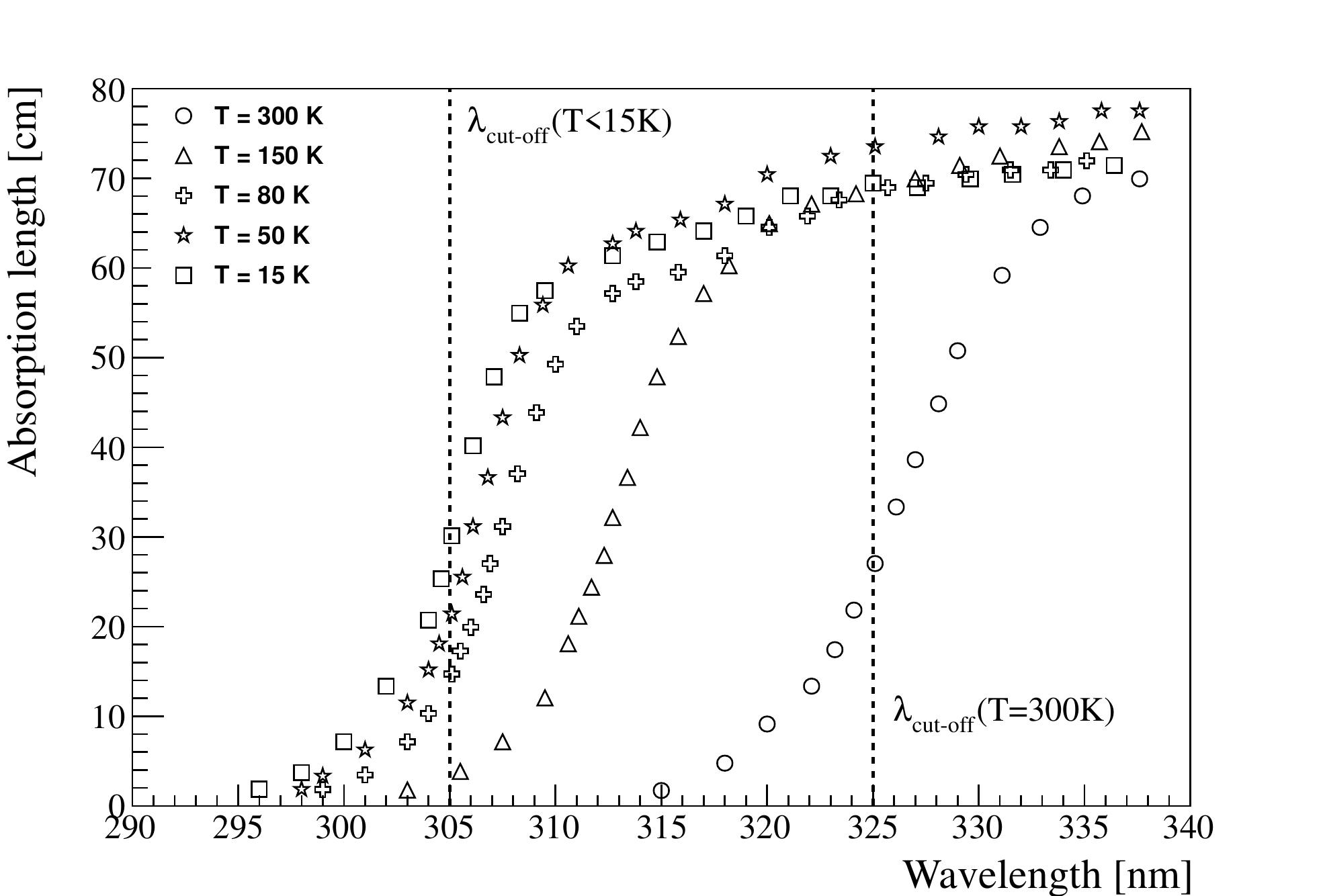}
 \caption{\label{fig:AbsLength} Absorption length for different temperatures around the fundamental absorption edge. The $\lambda_{cut-off}$ is defined as the inflection point of the curve. The reason why two $\lambda_{cut-off}$ values are shown in the plot is just to give a quantitative idea of the absorption spectrum variation. The simulation described below will not use these $\lambda_{cut-off}$ values but the entire absorption spectra.}
\end{figure}
We performed optical transmission measurements on pure {\TEO} slices using a Perkin Elmer Lambda 900 spectrophotometer and a Leybold RDK10-320 (T$_{\rm{min}}$=12~K) cryostat. The results around the fundamental absorption edge are shown in Fig.~\ref{fig:AbsLength}; for wavelengths longer than 340~nm the absorption length approaches a value of about 80~cm, irrespective of the temperature. The absorbance spectra show a temperature dependence typical of an indirect band structure, and the shift in the $\lambda_{cut-off}$ due to the temperature variation reaches an asymptotic value for T~$<50$~K. This suggests that for temperatures below 15~K it is possible to assume an absorbance spectrum like the one measured at 15~K.

In principle, also the temperature dependence of the refractive index may affect the number of Cherenkov photons but, as for the absorption length, no reference is available in literature. 
This systematic uncertainty was investigated by allowing the refracting index to vary according to the Lorentz theory for optical properties of dielectrics~\cite{wooten}, that foresees a maximum shift of the refractive index curve of about 20~nm toward lower wavelengths (as the variation is expected to follow the one of the absorption length). Using this value of the refractive index in Eq.~\ref{eq:NumberofPhotons} (instead of the room temperature refractive index), the number of Cherenkov photons emitted changes by less than 1\%. Therefore the temperature dependence of the refractive index will be neglected.


\subsection{Number of Cherenkov photons emitted per unit path length}
\label{subsec:NumberOfCherenkovPhotons}
Using the refractive index and the absorption length it is possible to evaluate the number and the wavelength spectrum of the emitted Cherenkov photons per unit path length of the charged particle.
First of all, an estimation of the Cherenkov threshold must be made: using, for example, the value of $n_{\rm{o}}(\lambda=400~\rm{nm}, T=300~K)=2.44$ the corresponding threshold for Cherenkov emission is about 50~keV for electrons and about 400~MeV for alpha particles ($\frac{v}{c}=\beta>\frac{1}{n}$).
\begin{figure}[htbp]
\begin{centering}
\includegraphics[width=.5\textwidth]{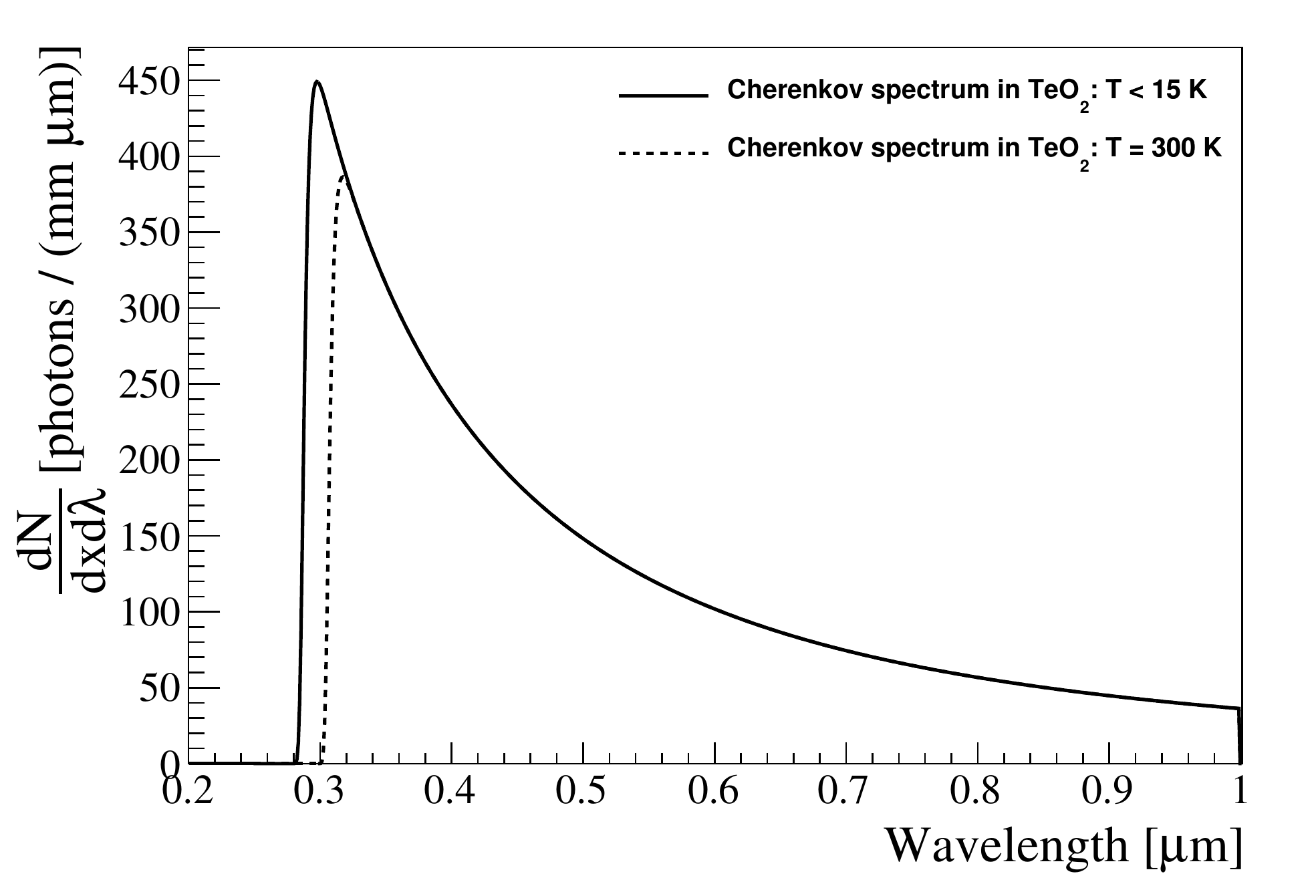}
\caption{\small{Emission spectrum of the Cherenkov photons in {\TEO} at room temperature and at cryogenic temperature (T~$<$~15~K). These spectra are evaluated using an electron of about 2.5~MeV for 1~mm path length.}}\label{fig:spectrumcherenkov}
\end{centering}
\end{figure}

In this example the number of Cherenkov photons produced by an electron of about 2.5~MeV for 1~mm path length is evaluated. Integrating Eq.~\ref{eq:NumberofPhotons} over $\lambda$ between 0.2 and 1.0~$\mu$m results in $84.1~\rm{photons/mm}$ for $n_{\rm{o}}$, and 86.5~photons/mm for $n_{\rm{e}}$.
At cryogenic temperature the $\lambda_{cut-off}$ decreases by about 20 nm; this effect increases the number of the emitted Cherenkov photons up to 92.9~photons/mm for $n_{\rm{o}}$, and 95.5~photons/mm for $n_{\rm{e}}$. The room temperature spectrum compared with the low temperature one is shown in Fig.~\ref{fig:spectrumcherenkov}, where it is also visible that the evaluation of the spectrum and number of Cherenkov photons was performed taking into account photons up to 1~$\mu$m. The near-infrared was chosen as end point of the wavelength spectrum of Cherenkov photons because most of the optical properties needed for the simulation of the photon propagation are available up to 1~$\mu$m. With this approximation the simulation takes into account 93\% of the total energy emitted by the Cherenkov effect (equivalent to $260~\rm{eV\cdot mm^{-1}}$) the remaining 7\% (equivalent to $20~\rm{eV\cdot mm^{-1}}$) is emitted at wavelength longer than 1~$\mu$m. This approximation is satisfactory for the purpose of the simulation presented in this paper. 

The {\TEO} stopping power for electrons can be found in~\cite{Nist}: thanks to this parameter it is possible to perform a raw approximation of the number and energy of Cherenkov photons produced by an electron with a kinetic energy of about 2.5~MeV. With a $dE/dx$ of about 0.8~MeV/mm its path results in 3~mm, corresponding to an average number of Cherenkov photons of $\sim280$ with a total energy of about 780~eV.

It should be mentioned that, even if the Cherenkov photons are emitted in a cone with an opening angle $\theta_{C} = \arccos(1 /\beta n)$ with respect to the electron's direction, they can be approximated (in our case) as isotropically emitted. The reason is that the electrons that emit Cherenkov light are affected by several direction changes due to multiple Coulomb scattering within the crystal. Furthermore, the surface of {\TEO} causes several reflections on the emitted photons. Thus, the initial directionality of the Cherenkov photons is lost and does not affect the mean end distribution of the Cherenkov energy detected, as confirmed by the measurements performed in Ref.~\cite{Beeman:2011yc, Casali:2014vvt}.

\section{Light trapping and surface effects of tellurium dioxide crystal}
\label{sec:LightTrapping}

The number of photons that escape the {\TEO} crystal is smaller than the 280 calculated above; this is due to the light trapping that happens when an optical photon travels from a medium with a higher refractive index ($n_1$) to one with a lower refractive index ($n_2$). If the incident angle of the photon on the crystal surface is grater than $\theta^{Critical}=\arcsin(n_2/n_1)$ the photon is totally reflected inside the crystal. For the {\TEO} surrounded by vacuum $\theta^{Critical}$ is shown in Fig.~\ref{fig:RIndex} for both the ordinary and the extraordinary refractive index. This small critical angle, in conjunction with to the high symmetry of the cubic shape of the crystal, leads to a high probability (about 75\%) that photons are internally reflected an indefinitely large number of times, i.e. they remain trapped inside the crystal until absorption.

Fortunately, this trapping effect can be partially overcome thanks to the roughness of the crystal surface and the birefringence of {\TEO}. Both these properties can change the reflection angle increasing the exit probability of the photons, as will be shown in Sec.~\ref{sec:roughness}.

\section{Comparison between Monte Carlo simulation and data}
\label{sec:Simulation}

As roughly evaluated in section~\ref{sec:CherenkovTheory}, the produced Cherenkov light for a 2.5~MeV electron amounts to about 780~eV, a very different value from the 100~eV measured in Ref.~\cite{Casali:2014vvt}. To understand this discrepancy we developed a detailed Monte Carlo simulation able to reproduce the Cherenkov photon production and propagation inside the experimental set-up. The simulation was split in two parts: (i) the total number of emitted Cherenkov photons was evaluated using Geant4, (ii) the propagation of these photons inside the different components of the experimental set-up was reproduced by means of Litrani~\cite{Gentit200235}, a general simulation software specifically developed for the propagation of optical photons.

\begin{figure}[htbp]
\begin{centering}
\includegraphics[width=.5\textwidth]{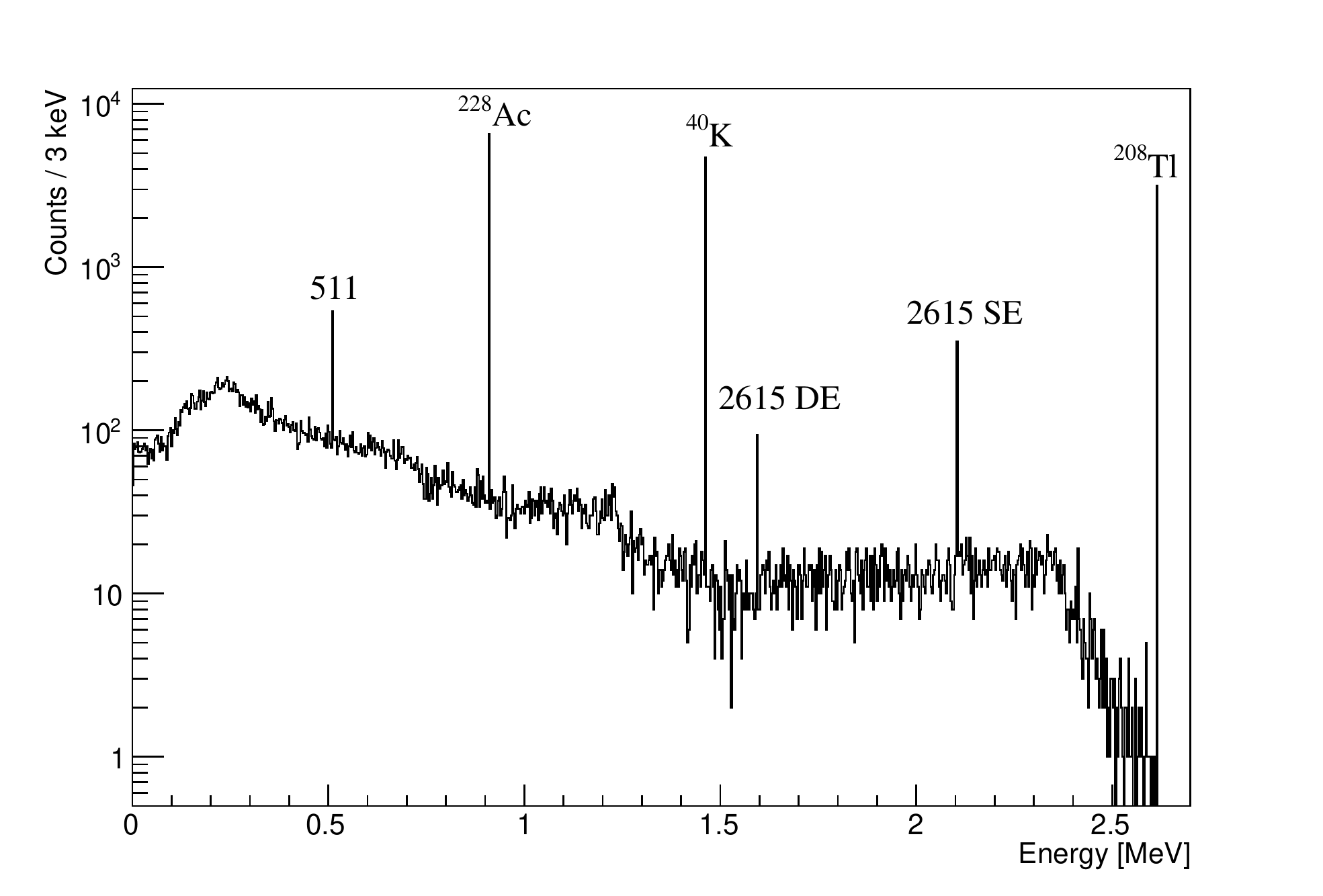}
\caption{\small{Spectrum of the energy deposited by the 2615~keV, 1460~keV and 911~keV ($^{208}$Tl, $^{40}$K and $^{228}$Ac respectively) $\gamma$s in the {\TEO} crystal as obtained from the Monte Carlo simulation (Geant4). In addition to the three photo-peaks we can see the SE, DE and the 511 keV photo-peak.}}\label{fig:spectrumTl}
\end{centering}
\end{figure}

The refractive index and the absorption length of the germanium in the optical wavelength range~\cite{PhysRevB.27.985}, as well as the real and imaginary part of the refractive index of the 3M VM2002 foil~\cite{Gentit200235} have been found in literature only at room temperature. Nevertheless, using the room temperature values does not affect the simulation results, as the reflectivity properties of materials are governed by Fresnel's equations, that depend only on their refractive indices. As shown for example in Ref~\cite{frey}, the refractive index of semiconductors (and in general of all the common materials) varies only by few percent in a wide temperature range, resulting in a variation of their reflectivity lower than 1\%. Similar remarks may be made about the absorbance capability of the germanium disk, because its thickness (300~$\mu$m) is orders of magnitude larger than its attenuation length (few nm as shown in Ref~\cite{PhysRevB.27.985}). Even if the attenuation length can slightly change at cryogenic temperature, its capability to stop an optical photon does not change. Given these considerations, we can state that approximating the optical properties of the germanium and of the reflective foil does not affect the simulation results.  
 
The surface roughness of the crystal is the only free parameter of the simulation. The reason is that the light diffusion produced by a rough surface is implemented in LITRANI exploiting the following approximation: the normal to the surface at the point hit by the photon is randomly tilted (with respect to the true normal of the surface) by an angle $\theta$. This angle is generated according to a distribution $sin(\theta)d\theta d\phi$, between 0 and $\theta_{rough}$. This parametrization is just an efficient Monte Carlo implementation of the more complex diffusion effect, therefore it cannot be compared to a direct measurement of the surface roughness, and it must be tuned from data. A detailed analysis of the light output as function of $\theta_{rough}$ is discussed in Sec.~\ref{sec:roughness}.

The Monte Carlo starts with the simulation of the three $\gamma$s that produce the six most intense peaks in the {\TEO} energy spectrum (see Ref.~\cite{Beeman:2011yc, Casali:2014vvt} for more details): 2615~keV, 1460~keV and 911~keV respectively emitted by $^{208}$Tl, $^{40}$K and $^{228}$Ac. The $\gamma$ emitted by $^{208}$Tl produces the 2615~keV photo-peak and two additional peaks: the single and double escape peaks (hereafter referred to as SE and DE), in which respectively one or two 511~keV $\gamma$ escapes from {\TEO} crystal after an $e^+ + e^-$ annihilation. If the $e^+ + e^-$ annihilation occurs in the inert materials or in a neighbour crystal, the 511~keV $\gamma$ can interact with the {\TEO} bolometer producing the 511~keV $\gamma$ photo-peak.

These three $\gamma$s are emitted by a thin radioactive wire placed just outside the cryostat, used as calibration source. A small fraction of them (2\%) interact with the detector setup and release their energy within the crystal. The total energy spectrum measured by {\TEO} after the propagation of $3\times10^6$ $\gamma$-rays is show in Fig.~\ref{fig:spectrumTl}.

\begin{figure}[htbp]
\begin{centering}
\includegraphics[width=.5\textwidth]{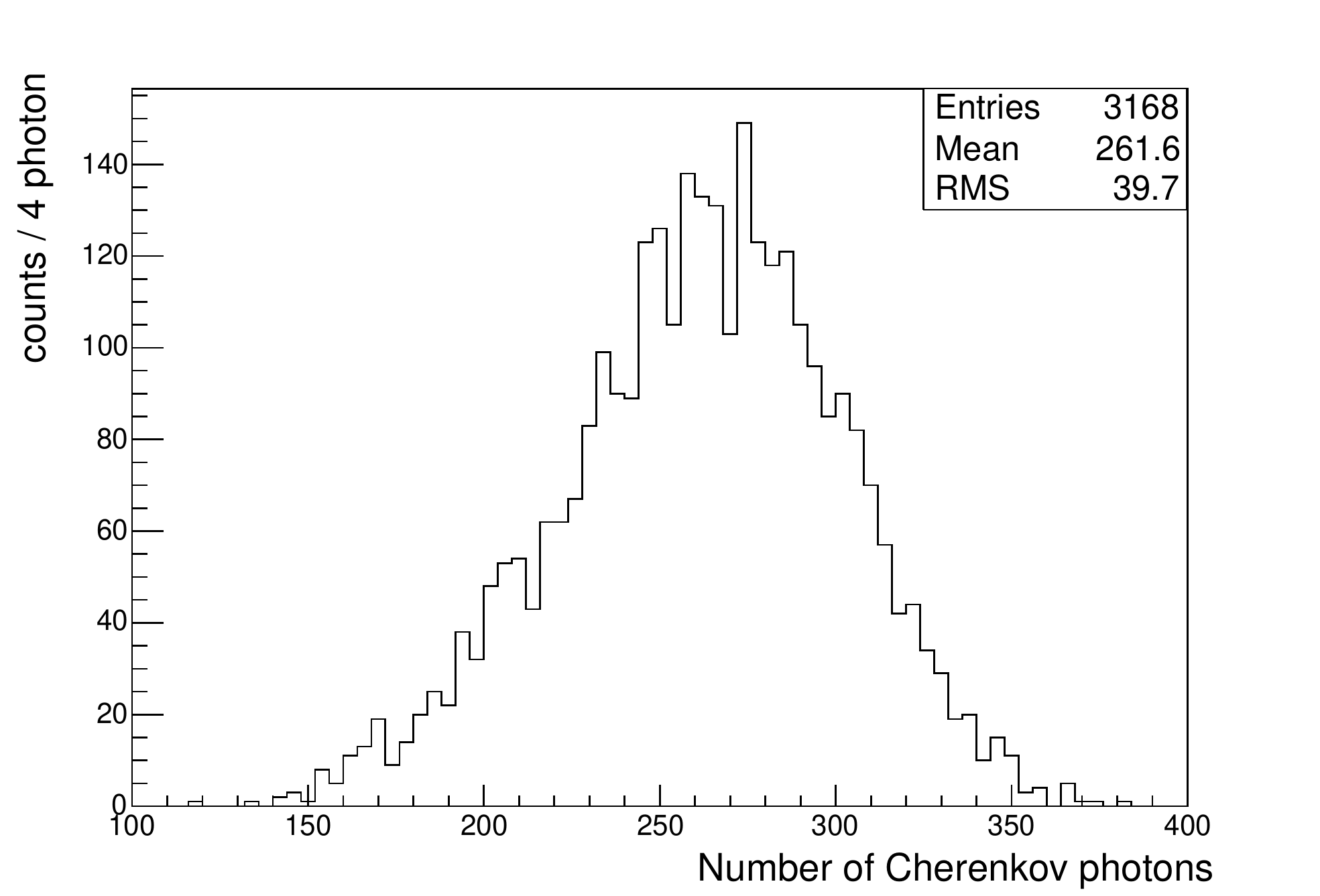}
\caption{\small{Distribution of Cherenkov photons produced by the 2615~keV $\gamma$ interactions that released all their energy in the crystal.}}\label{fig:ncherphot}
\end{centering}
\end{figure}

Selecting each photo-peak, it is possible to extract the kinetic energy and the path of all electrons and positrons produced in these interactions. Then, using equation~\ref{eq:NumberofPhotons}, we can assess the number of Cherenkov photons produced by each e$^{\pm}$ particle with kinetic energy greater than 50~keV, as shown in Fig.~\ref{fig:ncherphot} for the $^{208}$Tl photo-peak.
Once the number of Cherenkov photons produced is known, it is possible to simulate their propagation in the experimental set-up with Litrani; for this purpose also the optical properties of the germanium and 3M VM2002 reflective foil are needed~\cite{Gentit200235,PhysRevB.27.985}. 
For each photo-peak we simulate the light produced by 100 interactions. For each interaction, the total number of Cherenkov photons and their wavelength distribution are randomly generated respectively according to the distribution shown in Fig.~\ref{fig:ncherphot} and Fig.~\ref{fig:spectrumcherenkov}. In the particular case of $^{208}$Tl, the energy of the emitted Cherenkov photons results in a Gaussian with $\mu$ and $\sigma$ respectively of about 744~eV and 120~eV. The average energy is very close (less than 5\%) to the raw approximated value calculated in Sec.~\ref{sec:CherenkovTheory}.

\begin{figure}[htbp]
\begin{centering}
\includegraphics[width=.180\textwidth]{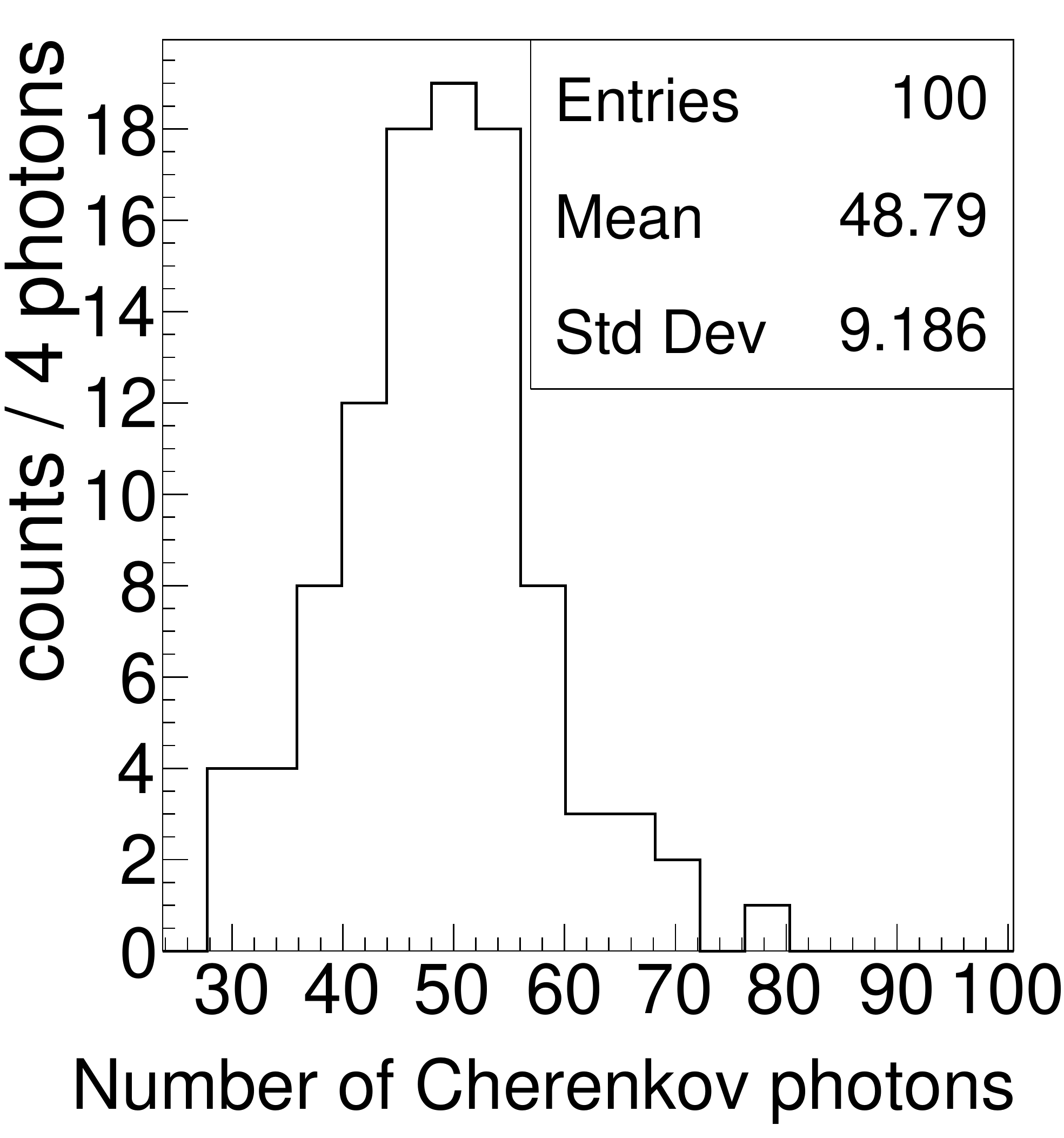}
\includegraphics[width=.280\textwidth]{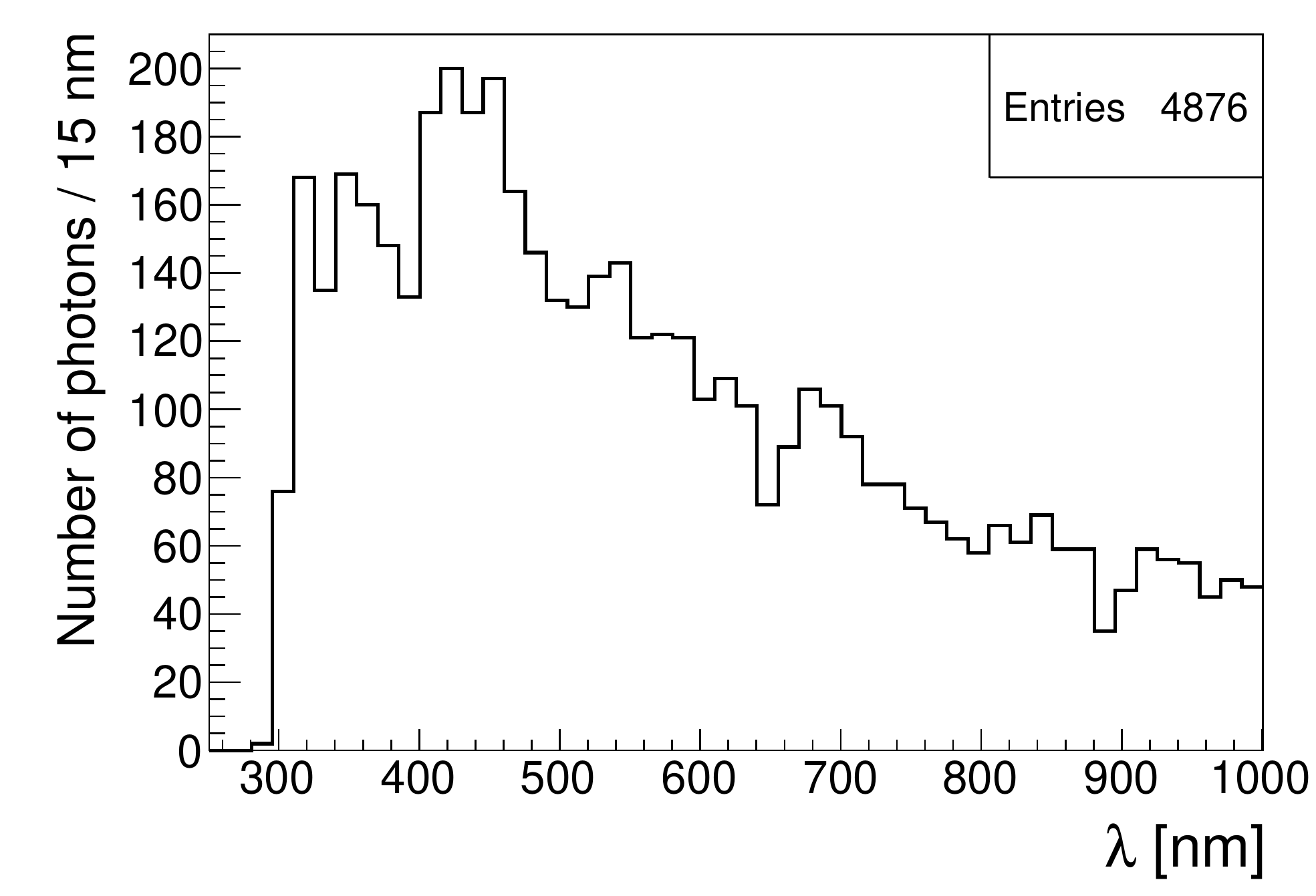}
\caption{\small{Distribution of the number of Cherenkov photons (left) and their wavelength distribution (right) absorbed by the germanium disk in one hundred 2615~keV $\gamma$-interactions simulated by means of Litrani. The reduction of detected photons between $300-400$~nm is due to the low reflectivity of the 3M VM2002 reflective foil in that wavelengths range.}}\label{fig:Cherephot2615Det}
\end{centering}
\end{figure}

Each Cherenkov photon is propagated through the materials until it is absorbed: if it is absorbed inside the germanium disk it is considered as detected. The number of detected photons and their wavelength distribution for the $^{208}$Tl peak are shown in Fig.~\ref{fig:Cherephot2615Det}.

\begin{figure}[htbp]
\begin{centering}
\includegraphics[width=.5\textwidth]{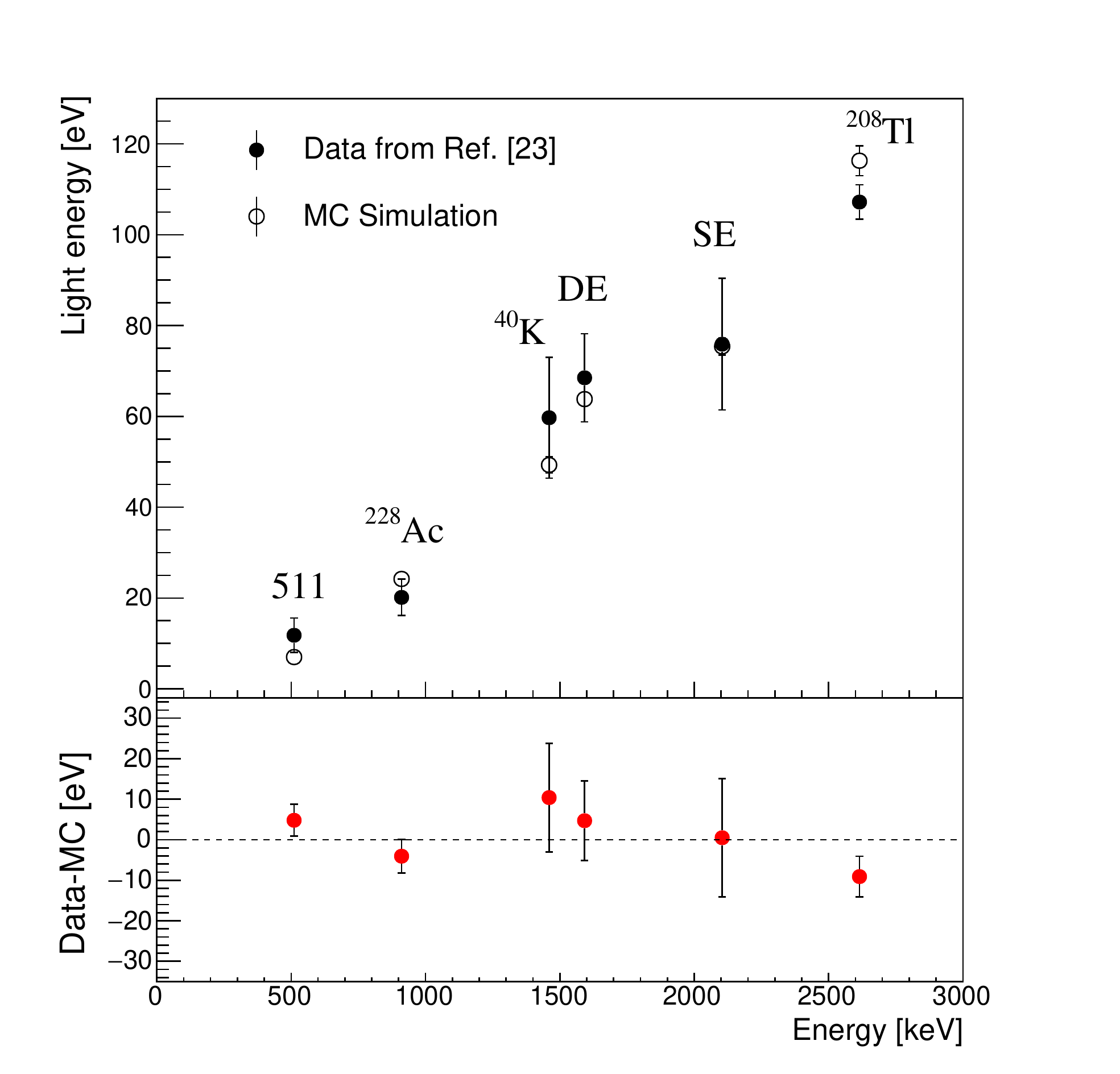}
\caption{\small{Top: energy release in the Ge light detector measured in Ref.~\cite{Casali:2014vvt} (solid black circles) and simulated (open black circles). The data and the simulation refer to a wrapped CUORE-like {\TEO} crystal faced to a germanium light detector (5~cm diameter, 300~$\mu$m thickness). Bottom: difference between data and Monte Carlo (red dots).}}\label{fig:simuvm2002}
\end{centering}
\end{figure}

The Cherenkov energy spectrum of the photons absorbed by the germanium light detector can be evaluated integrating event by event the spectra of Fig.~\ref{fig:Cherephot2615Det}. This spectrum results in a Gaussian with $\mu$ and $\sigma$ respectively of about 117~eV and 23~eV. The same procedure is applied to the other five peaks and the obtained energies are compared with those measured in Ref.~\cite{Casali:2014vvt} in Fig.~\ref{fig:simuvm2002}.

The simulation reproduces the experimental data and confirms that just 18\% of Cherenkov light is detected. This is because only 27\% of light is able to escape the unwrapped crystal face, and 69\% of it is absorbed by the Ge-disk. The remaining 82\% is trapped inside the detector set-up, and it is absorbed by the crystal itself (60\%) or by the reflective foil and the light detector copper frame (22\%) after many reflections (see Ref.~\cite{Casali:2014vvt} for details on the detector set-up).
The Monte Carlo simulation also confirms that no significant improvements in light collection is expected using other reflectors such as PTFE tape or aluminum foil.

In light of these considerations it is possible to understand why all the trials to increase the light collection efficiency were not successful: the trapping effect of the {\TEO} crystal and the reflectivity of the germanium disk do not allow to collect more than 18\% of the Cherenkov light produced by $\beta/\gamma$ interactions.




An additional cross-check was made to validate the simulation: the LY of the smaller {\TEO} crystal, described in Sec.~\ref{sec:introduction}, was entirely reproduced exploiting the Monte Carlo. In this case, the simulation predicts a higher LY with respect to the $5\times5\times5$~cm$^3$ crystal, due to the lower self-absorption. The detected light predicted by the simulation at the $^{208}$Tl line (2615~keV) is $205\pm4$~eV, in full agreement with the measured one ($195\pm7$~eV). This result can be considered as a further evidence that, despite the approximations, the Monte Carlo simulation developed in this work is able to reproduce and predict the Cherenkov light yield of {\TEO} bolometers.

The Monte Carlo results depend on the only free parameter of the simulation, $\theta_{rough}$, parameterizing the surface roughness of the crystal. A detailed study of the light emitted as function of $\theta_{rough}$ will be shown in the next sections.




\section{Effect of the crystal surfaces roughness on light yield}
\label{sec:roughness}
The only free parameter of the simulation, $\theta_{rough}$, is related to the surface roughness of the crystal. The surface treatment of the CUORE crystals was developed to maximize the surface radio-purity of the crystal. Due to {\TEO} hardness characteristics, the crystal features two opposite faces with a better polishing quality (close to optical polishing grade) and four faces with a higher roughness. Therefore, the crystal surface is not optimized for the light collection. In the measurements performed in Ref.~\cite{Beeman:2011yc,Casali:2014vvt}, the germanium light detector monitored one of the two polished faces of the crystal.

\begin{figure}[htbp]
\begin{centering}
\includegraphics[width=.5\textwidth]{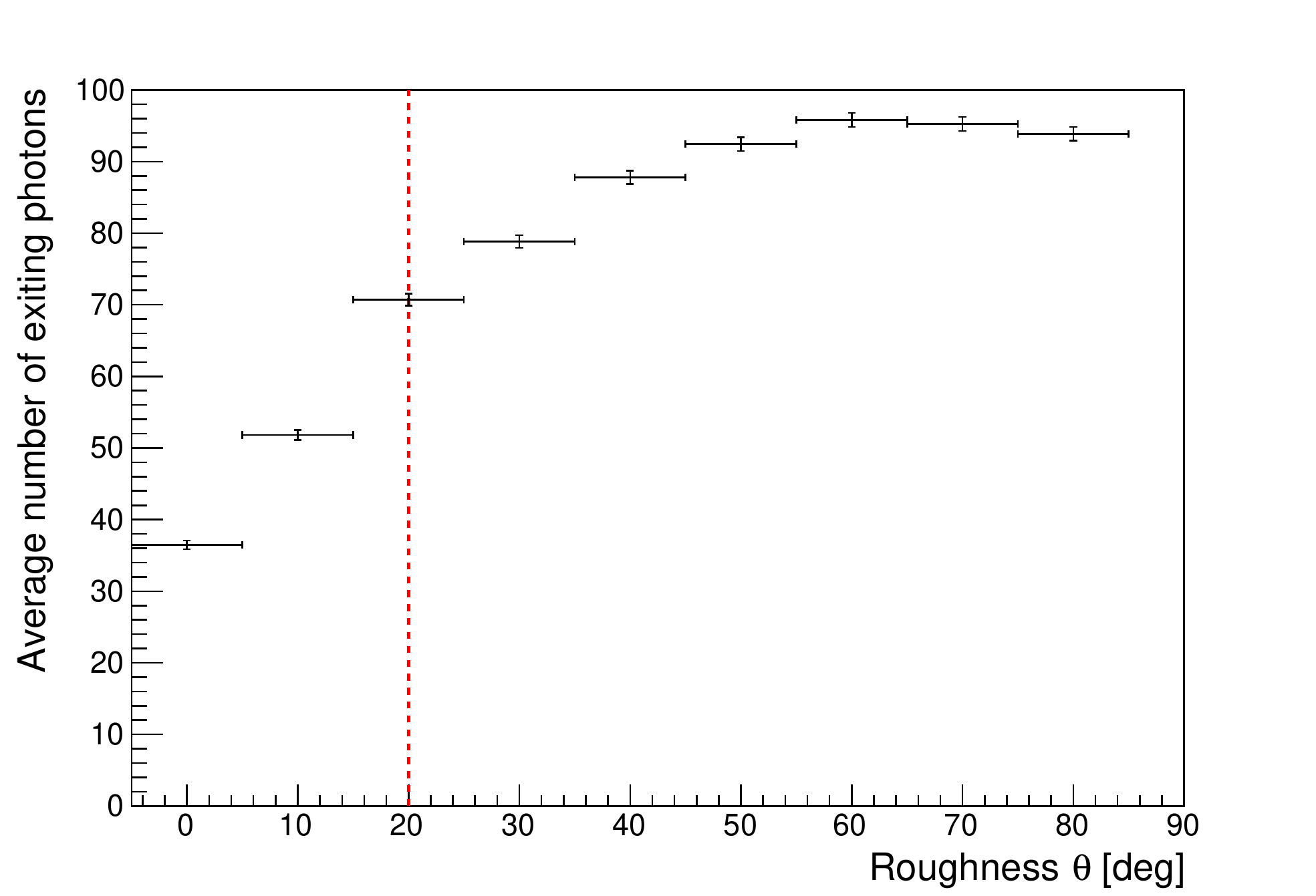}
\caption{\small{Average number of photons reaching the light detector (1.2~cm away from crystal) as a function of the roughness of the four lateral faces of the crystal.}}\label{fig:lightvsroug}
\end{centering}
\end{figure}

As discussed in Sec.~\ref{sec:LightTrapping}, the surface roughness interferes with the trapping effect because of the high refractive index and symmetry of the {\TEO} bolometers, increasing the photons emission probability. The result reported in Fig.~\ref{fig:lightvsroug} highlights the correlation between the emitted light and the surface roughness of the crystal faces. Fig.~\ref{fig:lightvsroug} also shows the value of $\theta_{rough}$ that reproduces all the cryogenic tests performed with cubic {\TEO} crystals (red dotted line $\theta_{rough}=20^{\circ}$).

The simulation suggests that the LY can be maximized by increasing the roughness of the four lateral faces of the crystal. The light yield can be increased by 35\% with respect to the standard CUORE {\TEO} crystals ($\theta_{rough}\sim20^{\circ}$) with a surface roughness of $\theta_{rough}\sim60^{\circ}$. No additional advantages are found assuming that also the two opposite polished faces of the {\TEO} crystal are roughened.

\section{Improving the light collection}
\label{sec:improve}

Also the shape of the germanium light detector can be optimized to maximize the light collection efficiency. The light detector of Ref.~\cite{Casali:2014vvt} was a 300~$\mu$m thick, 5~cm in diameter germanium disk. Since the best shape and size of the light detector is the one that matches the {\TEO} face, a $5\times5$~cm$^{2}$ and 300~$\mu$m thick germanium slab was simulated. The active surface of this new light detector is larger by 27\% with respect to the old one. The distance between the crystal face and the germanium slab does not affect significantly the light collection efficiency in the range between 1.2 and 0.2~cm (in the hypothesis that the reflective foil surrounds also the empty space between the crystal face and the light detector), and for simplicity it was set to 1~cm.
\begin{figure}[htbp]
\begin{centering}
\includegraphics[width=.5\textwidth]{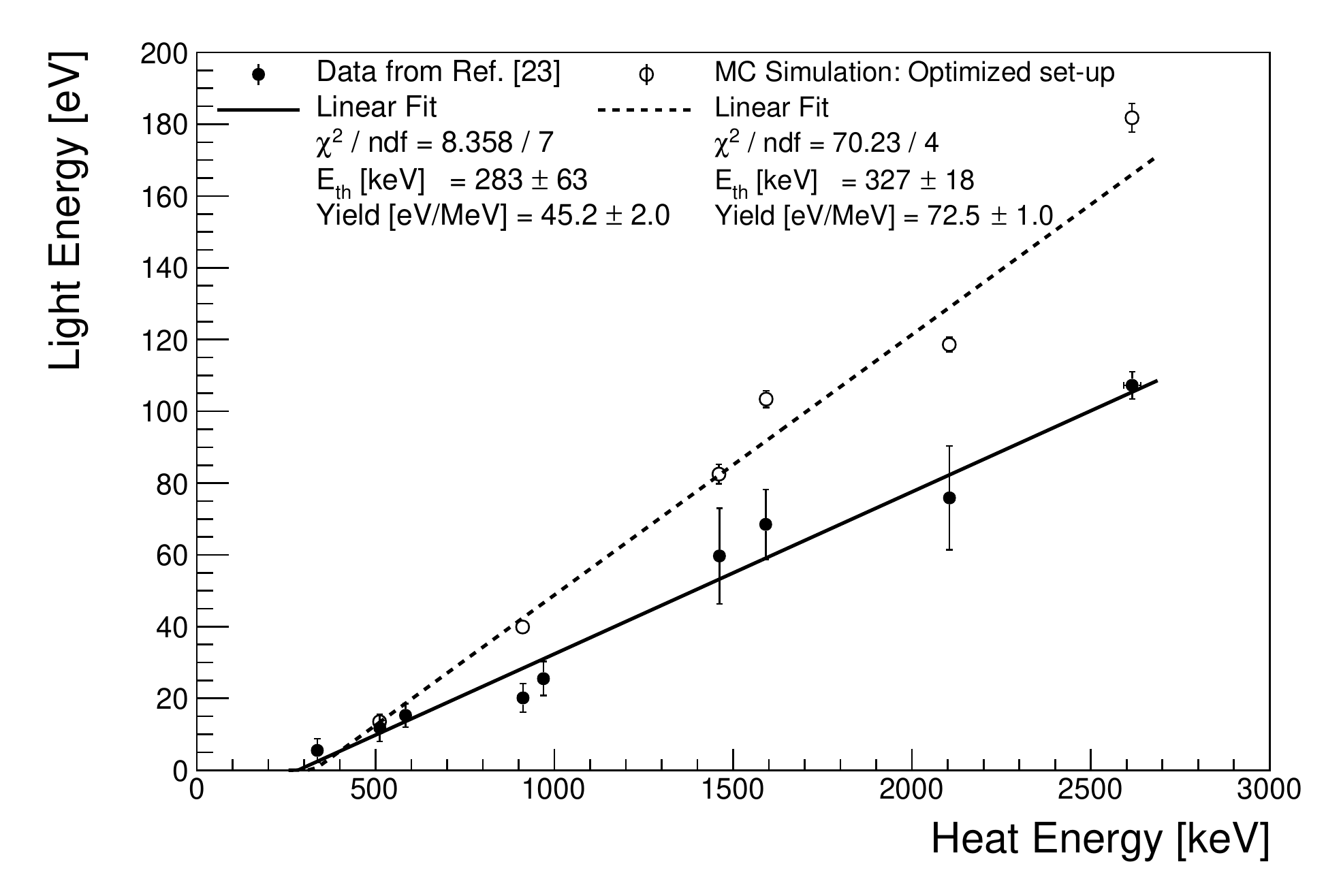}
\caption{\small{Comparison between the Cherenkov energy simulated in the optimized set-up as a function of the energy released in the {\TEO} bolometer (dotted line and open black circles) and the measured one~\cite{Casali:2014vvt} (black line and solid black circles). ${\rm E_{th}}$ is the energy value where the fit function crosses the Energy-axis. The high value of the fit $\chi^2$/ndf for the simulation can be explained by the fact that the linear fit is just a first approximation of the real trend of the Cherenkov light emission as a function of the energy of the $\gamma$ interactions, as explained in the text.}}\label{fig:sim_opt}
\end{centering}
\end{figure}
The mean values and the standard deviations of the light detected at the energies of the $\gamma$ peaks were derived by performing Gaussian fits (the same procedure adopted in section~\ref{sec:Simulation}). These data were fitted as a function of the energy with a first degree polinomial function (see Fig.~\ref{fig:sim_opt}). The fit shows that there is a minimum energy for the detection of Cherenkov light (amounting to ${\rm E_{th}} = 327\pm 18$~keV) and that the Cherenkov yield is $72\pm 1$~eV/MeV. The optimized set-up should be able to increase the {\TEO} Cherenkov yield by about $60\pm3\%$ with respect to the one measured in Ref.~\cite{Casali:2014vvt} in which the Cherenkov yield was $45.2\pm 2.0$~eV/MeV.


A remarkable improvement in light collection efficiency can be achieved also by improving the absorption of the Ge light detector. This can be done applying an antireflection coating on the Ge surface. Nevertheless, no reference is available in literature about a Ge antireflection coating that allows to increase the absorption over a wide range of wavelength (280-1000~nm) as well as over a wide range of incidence photons angle (0-90~degrees). Therefore this possibility was not investigated in this work.



\section{$\beta$, $\gamma$ and $\alpha$ discrimination}

Fig.~\ref{fig:sim_opt} shows that the linear function used in Ref.~\cite{Beeman:2011yc,Casali:2014vvt} to describe the Cherenkov light emission of {\TEO} bolometers is just a first approximation of the real trend of the Cherenkov light as a function of the $\beta/\gamma$ energy. For example, the events produced by the DE peak (1593~keV) are $e^-+e^+$ interactions that produced an average Cherenkov energy slightly greater than a $\gamma$ interaction with energy 1593~keV. Similarly, the events produced by the SE peak (2104~keV) are $e^-+e^++\gamma(511~\rm{keV})$. As expected, the Cherenkov light is equal to the one obtained for the DE peak plus the Cherenkov energy obtained for the 511~keV gamma interaction, and results slightly smaller than a $\gamma$ interaction with energy 2104~keV. A small deviation from the linear trend can be observed also for the 2615~keV peak.
\begin{figure}[htbp]
\begin{centering}
\includegraphics[width=.5\textwidth]{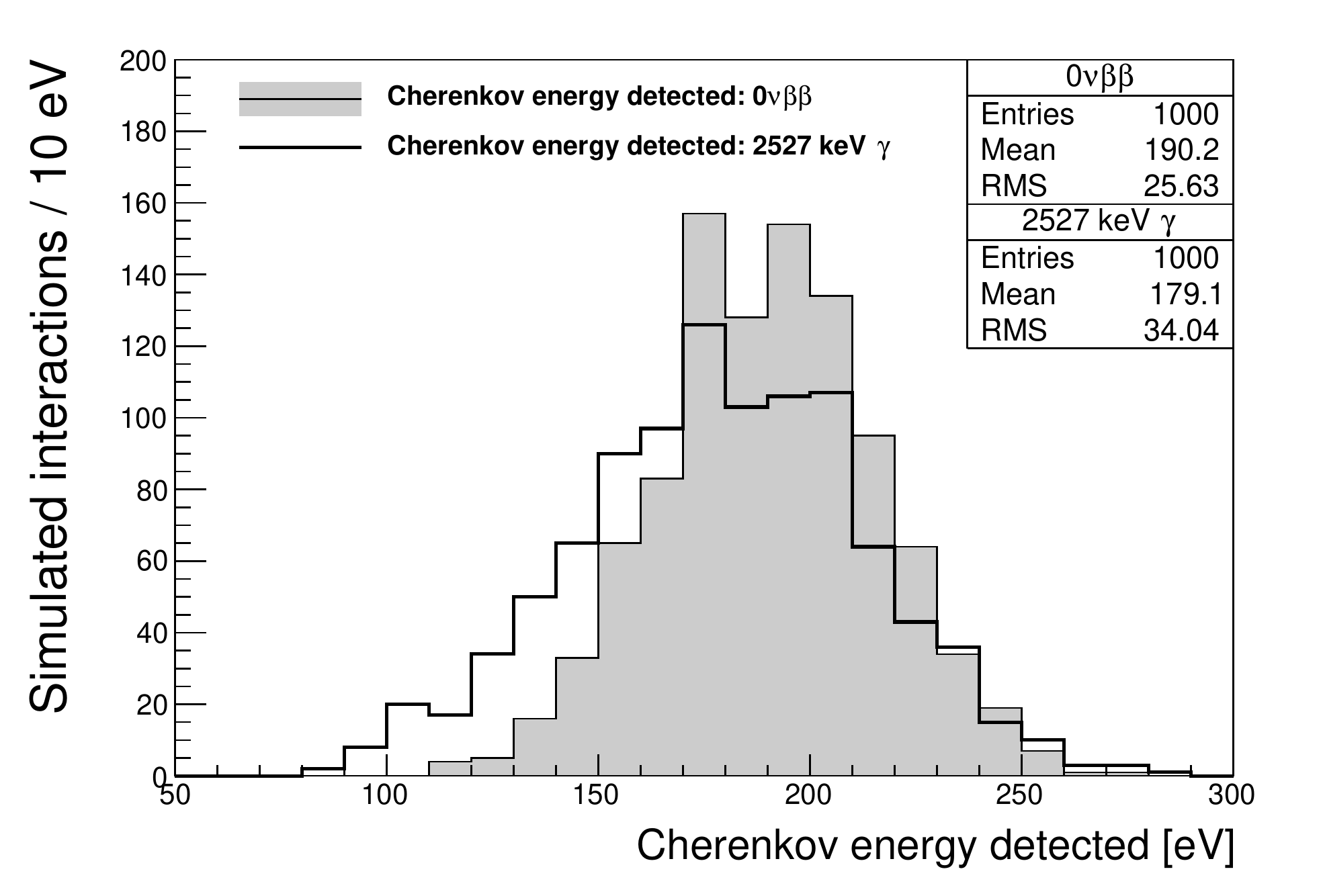}
\caption{\small{Energy distributions of the Cherenkov light produced by 1000 {\dbd} events of $^{130}$Te and 1000 $\gamma$ interactions with the same energy. The intrinsic fluctuations of these energy depositions are produced by the Poisson fluctuation of the number of emitted photons and by the dE/dx fluctuation along the electron's paths.}}\label{fig:betagammadiscr}
\end{centering}
\end{figure}

Given these considerations, it seems that $\beta$ and $\gamma$ interactions with the same energy result in a different Cherenkov yield. To study if also the $\gamma$ background can be removed from {\TEO} bolometers, the Cherenkov energy detected by the optimized set-up was simulated for {\dbd} events of $^{130}$Te, and then compared with the one produced by $\gamma$ interactions of the same energy.
The two energy distributions detected by the germanium slab are compared in Fig.~\ref{fig:betagammadiscr}. Even if the mean values are different, the distributions are partly overlapped, thus preventing the discrimination of a possible {\dbd} signal against the $\gamma$ background.

\begin{figure}[htbp]
\begin{centering}
\includegraphics[width=.5\textwidth]{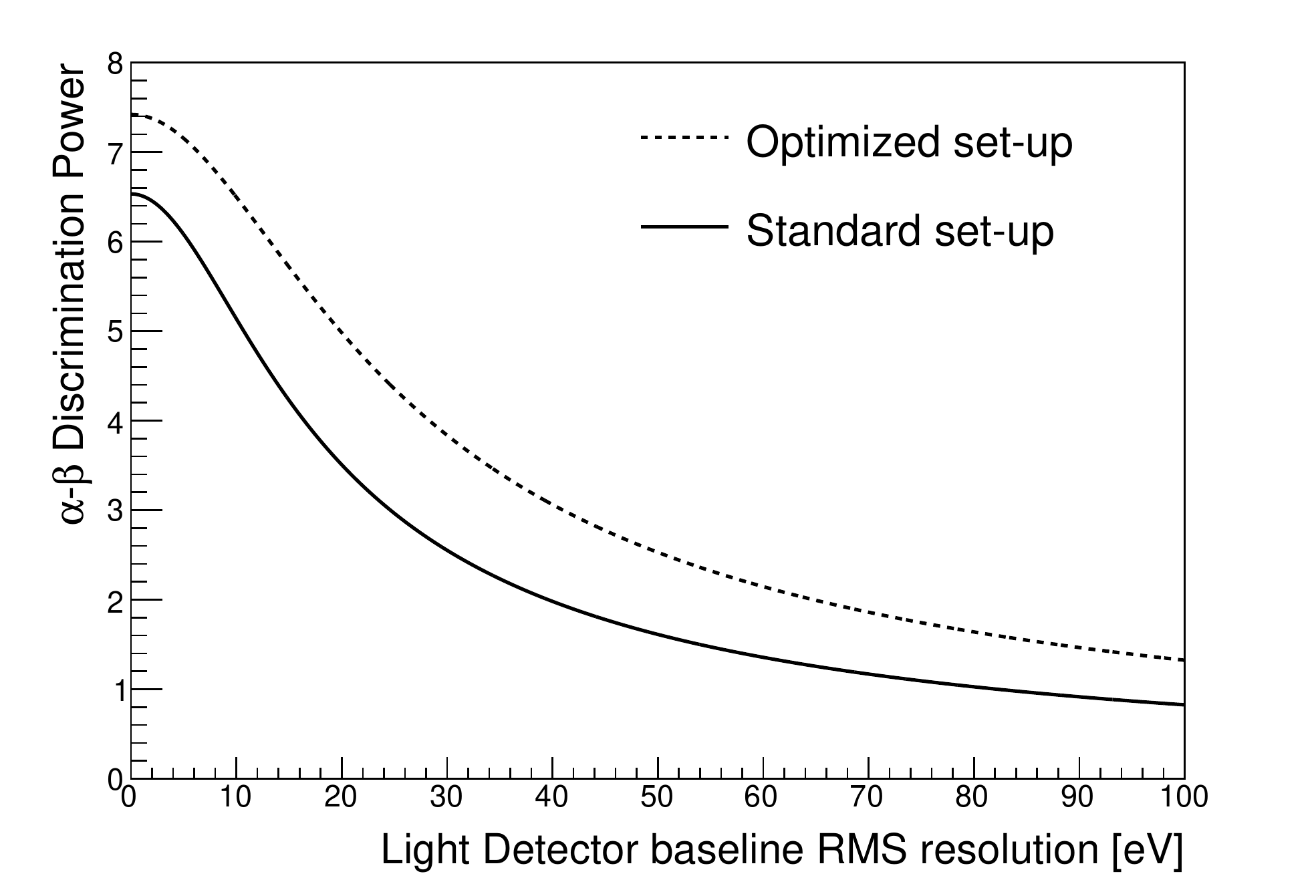}
\caption{\small{The discrimination power between $\alpha$ and $\beta$ interactions as a function of the baseline resolution of the light detector facing the {\TEO} bolometer for both the standard (continuous line) and the optimized (dotted line) set-up.}}\label{fig:dp}
\end{centering}
\end{figure}

On the contrary, the Cherenkov energy emitted by $\alpha$ particles with kinetic energy of few MeV is zero. Therefore, the discrimination power (DP) between these interactions and electrons is determined only by the light detector performance, i.e. its baseline resolution ($\sigma_{baseline}$):
\begin{equation}
DP= \frac{\mu_{\beta}}{\sqrt{2\sigma_{baseline}^2+\sigma_{\beta}^2}}\label{eq:dp}
\end{equation}
As stated in Ref.~\cite{Casali:2014vvt}, a very satisfactory $\alpha$ background rejection can be obtained for DP approaching 5. As shown in Fig.~\ref{fig:dp}, solving Eq.~\ref{eq:dp} for $\sigma_{baseline}$ and using the Cherenkov signal ($\mu_{\beta}$) and its intrinsic fluctuation ($\sigma_{\beta}$) estimated from the simulation, a baseline RMS resolution of 11~eV and 20~eV respectively for the standard and optimized setup was obtained.

Therefore, the rejection of the $\alpha$ background in a CUORE-like {\TEO} bolometer can be achieved exploiting light detectors with a baseline RMS energy resolution of about 11~eV. This constraint on the energy resolution of light detectors can be relaxed by a factor 2 (20 eV RMS) assuming that the light collection improvement simulated in Sec.~\ref{sec:improve} is actually feasible.

\section{Conclusions}
In this paper a complete simulation of the Cherenkov photon production and propagation in {\TEO} bolometers is presented. The results are in good agreement with the data presented in Ref.~\cite{Beeman:2011yc,Casali:2014vvt}. 
The simulation also suggests that the detectable Cherenkov light energy for a {\dbd} of $^{130}$Te can be increased up to 190~eV by increasing the roughness of the crystal and by optimizing the size and shape of germanium light detector. Also the possibility to perform a $\beta$-$\gamma$ discrimination was investigated, but the intrinsic fluctuation of the number of Cherenkov photons does not allow such kind of particle identification. On the contrary, a complete $\alpha$ background rejection can be achieved exploiting light detectors with a baseline RMS resolution below 20~eV (11~eV using the experimental set-up described in Refs.~\cite{Casali:2014vvt}).

\section*{Acknowledgements}
I would like to thank Dr. Ioan Dafinei for the great help in providing the transmission measurement of {\TEO} crystal and Dr. Marco Vignati for the continuous support offered in the realization of this work. I also would like to thank Dr. Laura Cardani for her careful and accurate revision of the manuscript. This work was supported by the Italian Ministry of Research under the FIRB contract no. RBFR1269SL and under the PRIN 2010-2011 contract no. 2010ZXAZK9.










\end{document}